\documentclass[aps,prd,twocolumn,nofootinbib]{revtex4-1}
\usepackage{epsfig}
\usepackage[colorlinks,linkcolor=blue,anchorcolor=blue,citecolor=blue,urlcolor=blue,breaklinks=true]{hyperref}
\usepackage{graphics}
\usepackage{color}
\usepackage{slashed}

\begin{document}

\author{Shu-Sheng Xu$^{1,2}$}~\email[]{Email: xuss@nju.edu.cn}

\author{Zhu-Fang Cui$^{1,2}$}~\email[]{Email: phycui@nju.edu.cn}

\author{An Sun$^{3}$}~\email[]{Email: sunan@nju.edu.cn}
\author{Hong-Shi Zong$^{1,2,4}$}~\email[]{Email: zonghs@nju.edu.cn}
\address{$^{1}$ Department of Physics, Nanjing University, Nanjing 210093, China}




\address{$^{2}$ State Key Laboratory of Theoretical Physics, Institute of Theoretical Physics, CAS, Beijing, 100190, China}

\address{$^{3}$College of Engineering and Applied Sciences, Nanjing University, Nanjing 210093,     China}
\address{$^{4}$ Joint Center for Particle, Nuclear Physics and Cosmology, Nanjing 210093, China}

\title{A new algorithm towards quasi-Wigner solution of the gap equation beyond the chiral limit}
\begin{abstract}
We propose a new algorithm to solve the quasi-Wigner solution of the gap equation beyond chiral limit. Employing a Gaussian gluon model and rainbow truncation, we find that the quasi-Wigner solution exists in a limited region of current quark mass, $m<43.1$~MeV, at zero temperature $T$ and zero chemical potential $\mu$.
 The difference between Cornwall-Jackiw-Tomboulis (CJT) effective actions of quasi-Wigner and Nambu-Goldstone solutions shows that the Nambu-Goldstone solution is chosen by physics.
Moreover, the quasi-Wigner solution is studied at finite temperature and chemical potential, the far infrared mass function of quasi-Wigner solution is negative and decrease along with $T$ at $\mu=0$. Its susceptibility is divergent at certain temperature with small $m$, and this temperature decreases along with $m$. Taking $T=80$~MeV as an example, the quasi-Wigner solution is shown at finite chemical potential upto $\mu=350$~MeV as well as Nambu solution, the coexistence of these two solutions indicates that the QCD system suffers the first order phase transition. The first order chiral phase transition line is determined by the difference of CJT effective actions.
\bigskip


\bigskip

\noindent PACS Number(s): 11.30.Rd, 25.75.Nq, 12.38.Mh, 12.39.-x

\end{abstract}
\maketitle

\section{Introduction}\label{intro}
Dynamical chiral symmetry breaking (DCSB) is one of the most important features of quantum chromodynamics (QCD), which plays significant role in low energy hadron physics. For instance, it gives a good explanation of the origin of constituent light quark masses and further plays key part for the mass of visible matter in the universe~\cite{national2013nuclear}. The mass function of the dressed quark propagator behaves a sharp increase around $p^2\lesssim 2~\mathrm{GeV}^2$ both in studies of Dyson-Schwinger equations (DSEs) and lattice QCD~\cite{PhysRevC.68.015203,PhysRevD.70.094039,ROBERTS200850,PhysRevD.66.014505,PhysRevD.70.034509,PhysRevD.76.094501}. It is believed that the chiral symmetry would be restored at sufficient high temperature, which indicates that there are two types of solutions of quark gap equation, namely Wigner solution for chiral symmetry phase and Nambu-Goldstone solution for DSCB.

The gap equation is nonlinear, which leads to the possibility of multiple solutions. In the chiral limit, it is well known that the perturbative method can not give a nonzero mass correction, and therefore the chiral symmetry solution (Wigner mode), which preserves vanishing quark condensate, exists when the interaction is weak. As the interaction turns stronger, a dynamical chiral symmetry breaking solution (Nambu-Goldstone mode) appears if the coupling constant is stronger than a critical value.
Analogously, one expects that there exists a Wigner-kindred solution, which we name it as ``quasi-Wigner'' solution~\footnote{In the chiral limit, the chiral symmetry is the strict symmetry in the original Lagrangian, while the chiral symmetry is slightly broken in the Lagrangian level in the case of small current quark mass. Therefore, ``quasi-Wigner'' solution means the effective mass of the dressed quark is small and comparable with the small current quarks mass.}, in the case of small current quark mass. Although the chiral symmetry is broken explicitly, the chiral symmetry is still approximately exist in the Lagrangian level because of the current quark mass is very small compared with the QCD intrinsic energy scale $\Lambda_{QCD}$. However, the quasi-Wigner solution can not be obtained by simple iterative numerical calculation as the case of Nambu-Goldstone solution.
In Ref.~\cite{Zong:2004nm}, the authors have given an example, albeit simple, to show the Wigner solution in the case of light current quark mass by series expansion method. From that time on, a lot of authors had made great efforts to study the Wigner solution and possible other solutions with various models and methods in the case of non-chiral limit~\cite{PhysRevC.75.015201, PhysRevD.86.114001, PhysRevLett.106.172301, Jiang:2012zzh, Cui:2013tva, Cui:2013aba}.
Especially, K. Wang \textit{et al}. employ two different interaction models and two vertex \textit{ans\"atz} and with the help of homotopy continuation method to study the multiple solutions of gap equation at zero temperature and zero chemical potential~\cite{PhysRevD.86.114001}. Furthermore, in Ref.~\cite{PhysRevLett.106.172301}, the authors have studied the Nambu-Goldstone and Wigner solutions and their susceptibilities simultaneously at finite temperature and chemical potential in the chiral limit, and thereafter determine the chiral phase transition lines.

Different from the above, herein we propose a new algorithm to solve the quasi-Wigner solution, which is based on the uniqueness of linear equations' solution. We focus on the existence of quasi-Wigner solution and its dependence of the current quark mass and temperature and further compare them with the case of Nambu-Goldstone solution. 

This paper is organized as follows. In Sec.~\ref{dseandalgorithm}, we give a brief introduction to gap equation of quark propagator and then describe the algorithm to solve quasi-Wigner solution. Sec.~\ref{wignersolution} discusses the quasi-Wigner solution varying with current quark mass at vacuum QCD and further applies this algorithm to the thermal and dense QCD system. 
With the help of the susceptibility and the CJT effective action, the details of the quasi-Wigner solution are discussed at finite $T$ and $\mu$.
Finally, the conclusions and a brief summary are given in Sec.~\ref{sum}.

\section{Gap equation and a new algorithm}\label{dseandalgorithm}
In quantum field theory, the correlation functions are basic elements and two point correlation functions, i.e. various propagators, are the most important components for phenomenological study. The DSEs are the equations of motion for these correlation functions. The non-perturbative DSEs' studies of two point correlation functions are extensively used in various topics, such as hadron physics~\cite{ROBERTS200850}, thermal and dense QCD sysems~\cite{Roberts:2000aa}, chiral phase transition in $2+1$ dimensional Quantum Electrodynamics ($\mathrm{QED}_3$)~\cite{Yin:2014zta,Yin:2016vml}. Herein, we briefly introduce gap equation and elaborate the algorithm of quasi-Wigner solution with finite current quark mass.
\subsection{Zero temperature and chemical potential}
The gap equation at zero temperature and chemical potential reads\footnote{We use a Euclidean metric in this study, that is $\{\gamma_\mu,\gamma_\nu\}=2\delta_{\mu\nu}$; $\gamma_\mu^\dagger=\gamma_\mu$; $a\cdot b=\sum_{i=1}^4 a_i b_i$.}
\begin{equation}
S(p)^{-1} = Z_2S_0(p)^{-1}(p) + \frac{4}{3} Z_{1F}\int_q^\Lambda g^2 D_{\mu\nu}(p-q)\gamma_\mu S(q)\Gamma_\nu,    \label{gapeq}
\end{equation}
where $S(p)$ is the dressed quark propagator, which can be generally expressed as three equivalent forms via structural analysis, namely
\begin{eqnarray}
S(p) &=& -i\sigma_v(p^2) \slashed{p} + \sigma_s(p^2)
\nonumber\\
&=& \frac{1}{iA(p^2)\slashed{p} + B(p^2)}
\nonumber\\
&=& \frac{Z(p^2)}{i\slashed{p} + M(p^2)},
\label{quarkp1}
\end{eqnarray}
and $S_0(p)$ is the bare quark propagator,
\begin{equation}
S_0(p) = \frac{1}{i\slashed{p} + Z_m m},
\label{quarkp2}
\end{equation}
$D_{\mu\nu}(p-q)$ is the dressed gluon propagator, $\Gamma_\nu$ is the one-particle-irreducible (1PI) quark-gluon vertex, $Z_2$, $Z_{1F}$ are field strength and quark-gluon-vertex renormalization constants respectively, and $\int_q^\Lambda := \int^\Lambda\frac{d^4q}{(2\pi)^4}$ is translational invariant integration with a sufficient large cutoff at $q^2=\Lambda^2$. Combining Eqs. (\ref{gapeq}), (\ref{quarkp1}) and (\ref{quarkp2}), one can get two coupled integral equations,
\begin{eqnarray}
A(p^2) &=& Z_2 + Z_{1F}\int_q^\Lambda \mathrm{tr}\left[ \frac{-i\slashed{p}}{4p^2} g^2 D_{\mu\nu}(p-q)\gamma_\mu S(q)\Gamma_\nu \right], \label{gapeqA}
\\
B(p^2) &=& Z_4 m + Z_{1F} \int_q^\Lambda \mathrm{tr}\left[ \frac{1}{4} g^2 D_{\mu\nu}(p-q) \gamma_\mu S(q) \Gamma_\nu \right],   \label{gapeqB}
\end{eqnarray}
where $Z_2$ and $Z_4=Z_2 Z_m$ are renormalization constants, which are determined by renormalization condition
\begin{eqnarray}
A(\zeta^2) &=& 1,   \label{renormA}
\\
B(\zeta^2) &=& m(\zeta^2),  \label{renormB}
\end{eqnarray}
here $\zeta$ is a renormalization energy scale, which is usually chosen at ultraviolet region. Combining Eqs. (\ref{gapeqA}), (\ref{gapeqB}), (\ref{renormA}) and (\ref{renormB}), one obtains
\begin{eqnarray}
Z_2 &=& 1 - Z_{1F}\int_q^\Lambda \mathrm{tr}\left[ \frac{-i\slashed{p}}{4p^2} g^2 D_{\mu\nu}(p-q)\gamma_\mu S(q)\Gamma_\nu \right]_{p^2=\zeta^2}, \label{z2}
\\
Z_4 m &=& m - Z_{1F} \int_q^\Lambda \mathrm{tr}\left[ \frac{1}{4} g^2 D_{\mu\nu}(p-q) \gamma_\mu S(q) \Gamma_\nu \right]_{p^=\zeta^2}.   \label{z4}
\end{eqnarray}

The gap equation can be solved if some truncations are employed and the dressed gluon propagator and 1PI quark-gluon-vertex are specified so that the Eqs. (\ref{gapeqA}) and (\ref{gapeqB}) are closed. The iterative method are commonly used to solve this equation due to its nonlinearity. In the case of chiral limit, the chiral symmetry breaking solution occurs if the strength of interaction larger than a critical value. The chiral symmetry preserving solution, that is Wigner solution, can be obtained by setting $B(p^2)=0$ as its solution and merely solve the Eq. (\ref{gapeqA}). However, such trick is invalid in the case of non-zero current quark mass since we do not know $B(p^2)$ of quasi-Wigner solution in advance, acually we even do not know whether it is exist or not.

Our algorithm of quasi-Wigner solution is based on the Wigner solution in the chiral limit and the derivative equation of Eqs. (\ref{gapeqA}) and (\ref{gapeqB}). The current quark mass $m$ is viewed as a parameter of gap equation, it could be regarded as a variable of quark propagator, which implies that $S(p,m)^{-1}=iA(p^2,m)\slashed{p} + B(p^2,m)$. Assuming that the Wigner solution in the chiral limit has already known, i.e. $A_W(p^2,0)$ and $B_W(p^2,0)=0$. One could obtain the equations of $A^\prime(p^2,m)$ and $B^\prime(p^2,m)$ by means of the derivative of Eqs. (\ref{gapeqA}) and (\ref{gapeqB}) with respect to $m$,
\begin{eqnarray}
A^\prime(p^2,m) &:=& \frac{\partial A(p^2,m)}{\partial m} = \frac{\partial\mathcal{F}^a(A,B)}{\partial m}, \label{gapeqAm}
\\
B^\prime(p^2,m) &:=& \frac{\partial B(p^2,m)}{\partial m} = 1 + \frac{\partial\mathcal{F}^b(A,B)}{\partial m} \label{gapeqBm},
\end{eqnarray}
where
\begin{eqnarray}
\mathcal{F}^a(A,B) &=& Z_{1F}\int_q^\Lambda \mathrm{tr}\left[ \frac{-i\slashed{p}}{4p^2} g^2 D_{\mu\nu}(p-q)\gamma_\mu S(q)\Gamma_\nu \right]
\nonumber\\
&&-Z_{1F}\int_q^\Lambda \mathrm{tr}\left[ \frac{-i\slashed{p}}{4p^2} g^2 D_{\mu\nu}(p-q)\gamma_\mu S(q)\Gamma_\nu \right]_{p^2=\zeta^2},
\\
\mathcal{F}^b(A,B) &=& Z_{1F}\int_q^\Lambda \mathrm{tr}\left[ \frac{1}{4} g^2 D_{\mu\nu}(p-q)\gamma_\mu S(q)\Gamma_\nu \right]
\nonumber\\
&&-Z_{1F}\int_q^\Lambda \mathrm{tr}\left[ \frac{1}{4} g^2 D_{\mu\nu}(p-q)\gamma_\mu S(q)\Gamma_\nu \right]_{p^2=\zeta^2}.
\end{eqnarray}
Because of the closure of Eqs. (\ref{gapeqA}) and (\ref{gapeqB}), the integrations in the right-hand side of Eqs. (\ref{gapeqAm}) and (\ref{gapeqBm}) are $A(p^2,m)$ and $B(p^2,m)$ dependence, which are only unknown elements in advance. With the help of the identity
\begin{equation}
\frac{\partial\mathcal{F}(f)}{\partial \alpha} = \int dx \frac{\delta\mathcal{F}(f)}{\delta f(x)} \frac{\partial f(x)}{\partial \alpha},
\end{equation}
where $f$ is a normal function, $\mathcal{F}$ is a functional of $f$, and $\delta$ is variation operator, the Eqs. (\ref{gapeqAm}) and (\ref{gapeqBm}) can be read as
\begin{eqnarray}
A^\prime(p^2,m) &=& \int dk^2 \Big(\frac{\delta\mathcal{F}^a(A,B)}{\delta A(k^2)} A^\prime(k^2,m) 
\nonumber\\
&&\hspace*{10mm}
+ \frac{\delta\mathcal{F}^a(A,B)}{\delta B(k^2)} B^\prime(k^2,m) \Big),   \label{gapeqAm2}
\\
B^\prime(p^2,m) &=& 1+\int dk^2 \Big(\frac{\delta\mathcal{F}^b(A,B)}{\delta A(k^2)} A^\prime(k^2,m) 
\nonumber\\
&&\hspace*{10mm}
+ \frac{\delta\mathcal{F}^b(A,B)}{\delta B(k^2)} B^\prime(k^2,m) \Big).   \label{gapeqBm2}
\end{eqnarray}
We get a coupled linear equations for $A^\prime(p^2,m)$ and $B^\prime(p^2,m)$, which have only one solution if the matrix of $\frac{\delta\mathcal{F}^{\{a,b\}}(A,B)}{\delta \{A(k^2),B(k^2)\}}$ is not singular. Therefore, the quasi-Wigner solution of gap equation can be obtained with the help of $A$, $B$ functions in the case of chiral limit and the functions of $A^\prime$ and $B^\prime$, that is
\begin{eqnarray}
A_W(p^2, m) &=& A_W(p^2, 0) + \int_0^m d\bar{m} A_W'(p^2,\bar{m}),
\\
B_W(p^2, m) &=& B_W(p^2, 0) + \int_0^m d\bar{m} B_W'(p^2,\bar{m}).
\end{eqnarray}
Discretize the range of $\bar m$, assuming grids of $N$ are chosen as $\Delta m=\frac{m}{N}$, the quasi-Wigner solution for small mass $\Delta m$ is
\begin{eqnarray}
A_W(p^2,\Delta m) &=& A_W(p^2,0) + A^\prime_W(p^2,0)\Delta m,
\\
B_W(p^2,\Delta m) &=& B_W(p^2,0) + B^\prime_W(p^2,0)\Delta m
\nonumber\\
&=&B^\prime_W(p^2,0)\Delta m,
\end{eqnarray}
and for mass $m=n\Delta m$,
\begin{eqnarray}
A_W(p^2,(n+1)\Delta m) &=& A_W(p^2,n\Delta m) + A^\prime_W(p^2,n\Delta m)\Delta m,
\\
B_W(p^2,(n+1)\Delta m) &=& B_W(p^2,n\Delta m) + B^\prime_W(p^2,n\Delta m)\Delta m.
\end{eqnarray}
This is just the discretization version of integration. By the way, this algorithm can also be employed in solving Nambu-Goldstone solution, which tested its validity from another side.
\subsection{Finite temperature and finite chemical potential}
The chiral symmetry partially restored phase, corresponding quasi-Wigner solution of gap equation, is expected to occur at high temperature and (or) high chemical potential. 
Herein we describe the new algorithm applying to the gap equation at finite $T$ and $\mu$:
\begin{eqnarray}
S(\vec{p},\tilde{\omega}_n)^{-1} &=& Z_2^A i\slashed{\vec{p}} + Z_2i\gamma_4\tilde{\omega}_n + Z_4 m
\nonumber\\
&&+ Z_{1F}\sum_{\ell,\vec{q}}\!\!\!\!\!\!\!\!\int g^2 D_{\mu\nu}(\vec{k},\Omega_{n\ell}) \gamma_\mu S(\vec{q},\tilde{\omega}_\ell) \Gamma_\nu,    \label{gapeqT}
\end{eqnarray}
where $\slashed{\vec{p}}=\vec{\gamma}\cdot\vec{p}$, $\vec{k} = \vec{p}-\vec{q}$, $\tilde{\omega}_n=\omega_n+i\mu$ and $\omega_n=(2n+1)\pi T$ is Matsubara frequencies, $\Omega_{n\ell}=2(n-\ell)\pi T$,
\begin{equation}
\sum_{\ell,\vec{q}}\!\!\!\!\!\!\!\!\int := T\sum_{\ell=-\infty}^{+\infty} \int\frac{d^3\vec{q}}{(2\pi)^3},
\end{equation}
$Z_2^A, Z_2$ and $Z_4$ are renormalization constants, they are confirmed by renormalization condition
\begin{equation}
S(\vec{p},\omega_n)^{-1}|_{\vec{p}^2+\omega_n^2=\zeta^2} = i\slashed{\vec{p}} + i\gamma_4\omega_n + m.    \label{renormT}
\end{equation}
Because of nonzero $T$ and $\mu$ breaks the $O(4)$ symmetry of QCD systems into $O(3)$, the Dirac structures of the inverse of the dressed quark propagator at finite $T$ and $\mu$ become
\begin{eqnarray}
S(\vec{p},\tilde{\omega}_n)^{-1} &=&  \sum_{i=1}^4 \tau_i(\vec{p},\tilde{\omega}_n)A_i(\vec{p}^2,\tilde{\omega}_n),    \label{Sstruc}
\end{eqnarray}
with
\begin{eqnarray}
\tau_1(\vec{p},\tilde{\omega}_n) &=& i\slashed{\vec{p}},
\hspace*{10mm}
\tau_2(\vec{p},\tilde{\omega}_n) = \mathbf{I},
\nonumber\\
\tau_3(\vec{p},\tilde{\omega}_n) &=& i\gamma_4\tilde{\omega}_n,
\hspace*{4.5mm}
\tau_4(\vec{p},\tilde{\omega}_n) = \slashed{\vec{p}}\gamma_4\tilde{\omega}_n.
\end{eqnarray}
and $A_i(\vec{p}^2,\tilde{\omega}_n), (i=1,2,3,4)$, are 4 scalar functions.
Combining Eqs. (\ref{gapeqT}), (\ref{renormT}) and (\ref{Sstruc}),
\begin{eqnarray}
A_i(\vec{p}^2,\tilde{\omega}_n) &=& a_i + \mathcal{F}^i(A_1,A_2,A_3,A_4),
\end{eqnarray}
with
\begin{eqnarray}
&&\mathcal{F}^i(A_1,A_2,A_3,A_4) 
\nonumber\\
&=& Z_{1F}\sum_{\ell,\vec{q}}\!\!\!\!\!\!\!\!\int \mathrm{tr}\left[\bar\tau_i g^2 D_{\mu\nu}(\vec{k},\Omega_{n\ell}) \gamma_\mu S(\vec{q},\tilde{\omega}_\ell) \Gamma_\nu \right]
\nonumber\\
&-& Z_{1F}\sum_{\ell,\vec{q}}\!\!\!\!\!\!\!\!\int \mathrm{tr}\left[\bar\tau_i g^2 D_{\mu\nu}(\vec{k},\Omega_{n\ell}) \gamma_\mu S(\vec{q},\tilde{\omega}_\ell) \Gamma_\nu \right]_{\vec{p}^2+\omega_n^2=\zeta^2},
\end{eqnarray}
where $i=1,2,3,4$, $a_1=1, a_2=m, a_3=1, a_4=0$, and
\begin{eqnarray}
\bar\tau_1 &=& \frac{-i\slashed{\vec{p}}}{4\vec{p}^2},
\hspace*{10mm}
\bar\tau_2 = \frac{1}{4},
\nonumber\\
\bar\tau_3 &=& \frac{-i\gamma_4}{4\tilde{\omega}_n},
\hspace*{8.5mm}
\bar\tau_4 = \frac{\gamma_4\slashed{\vec{p}}}{4\vec{p}^2\tilde{\omega}_n}.
\end{eqnarray}
Using the same technique of zero temperature and chemical potential,
\begin{eqnarray}
A_i'(\vec{p}^2,\tilde{\omega}_n) &=& a_i' + \sum_{i\ell\vec{k}}\!\!\!\!\!\!\!\!\int \frac{\delta\mathcal{F}^i(A,B,C,D)}{\delta A_i(\vec{k}^2,\tilde{\omega}_\ell)} A_i'(\vec{k}^2,\tilde{\omega}_\ell),
\end{eqnarray}
where $a_1'=a_3'=a_4'=0$, $a_2'=1$.
Regarding the current quark mass $m$ as a variable of functions $A_i, (i=1,2,3,4)$, $A_2(\vec{p}^2,\tilde{\omega}_n)$ is known for Wigner solution in the chiral limit, that is $A_{W_2}(\vec{p}^2,\tilde{\omega}_n,0)=0$. Then, one can iterative the remain three equations to obtain the Wigner solution in the chiral limit. Similar with the case of $T=0=\mu$,
\begin{equation}
A_{Wi}(\vec{p}^2,\tilde{\omega}_n,m) = A_{Wi}(\vec{p}^2,\tilde{\omega}_n,0) + \int_0^m d\bar{m} A_{Wi}'(\vec{p}^2,\tilde{\omega}_n,\bar{m}).
\end{equation}
Discretizing the range of $\bar{m}$, $\Delta m=\frac{m}{N}$, then
\begin{equation}
A_{Wi}(\vec{p}^2,\tilde{\omega}_n,\Delta m) = A_{Wi}(\vec{p}^2,\tilde{\omega}_n,0) + A_{Wi}'(\vec{p}^2,\tilde{\omega}_n,0) \Delta m,
\end{equation}
and for general number $k$,
\begin{eqnarray}
&&A_{Wi}(\vec{p}^2,\tilde{\omega}_n,(k+1)\Delta m) 
\nonumber\\
&=& A_{Wi}(\vec{p}^2,\tilde{\omega}_n,k\Delta m) + A_{Wi}'(\vec{p}^2,\tilde{\omega}_n,k\Delta m) \Delta m.
\end{eqnarray}
One could get the quasi-Wigner solution in this way in its analytic region of current quark mass. In the next section, we will employ such algorithm into concrete model to show how the quasi-Wigner solutions varying with $m$ and $T$.
\section{Wigner solution and QCD phase diagram}\label{wignersolution}
In order to show some numerical results, we employ a consistent truncation and \textit{ans\"atz} for both of zero and non-zero temperature. The rainbow truncation
\begin{equation}
Z_{1F}\Gamma_\nu(q,p) = \gamma_\nu,
\end{equation}
which is widely used in studies of hadron physics and QCD phase diagram~\cite{PhysRevD.94.094030,PhysRevD.94.076009,PhysRevC.60.055214,Maris:1997hd,PhysRevC.56.3369,Xu:2015vna}.
and making the replacement
\begin{equation}
g^2D_{\mu\nu}(k) \longrightarrow \mathcal{G}(k) {D_0}_{\mu\nu}(k)
\end{equation}
in Eqs. (\ref{gapeq}) and (\ref{gapeqT}). ${D_0}_{\mu\nu}(k) = \delta_{\mu\nu} - \frac{k_\mu k_\nu}{k^2}$ is the free gluon propagator in Landau gauge, and $\mathcal{G}(k)$ is the effective interaction model, the form in this work is
\begin{equation}
\mathcal{G}(k^2) = D\frac{4\pi^2}{\sigma^6} k^2 e^{-\frac{k^2}{\sigma^2}},
\end{equation}
which is a ultraviolet dumping model and the renormalization is not necessary, $D$ and $\sigma$ are two parameters, which are usually fixed by observables in hadron physics. We use typical values, that is $\sigma=0.4$, $D\sigma = (0.72~\mathrm{GeV})^3$~\cite{Maris:2002mt}. At finite temperature, all $k^2$ are replaced by $\vec{k}^2 + \Omega_{n\ell}^2$.
\subsection{zero temperature}\label{zeroT}
\begin{figure}
\includegraphics[width=0.5\textwidth]{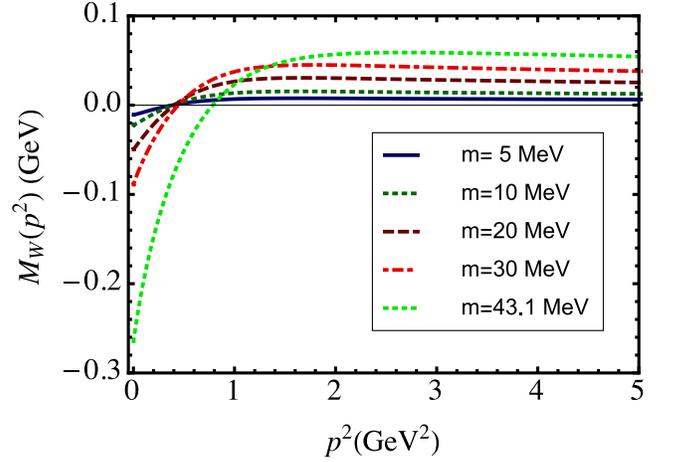}
\caption{The mass functions of Wigner solution with different current quark masses.}
\label{fig1}
\end{figure}
\begin{figure}
\includegraphics[width=0.5\textwidth]{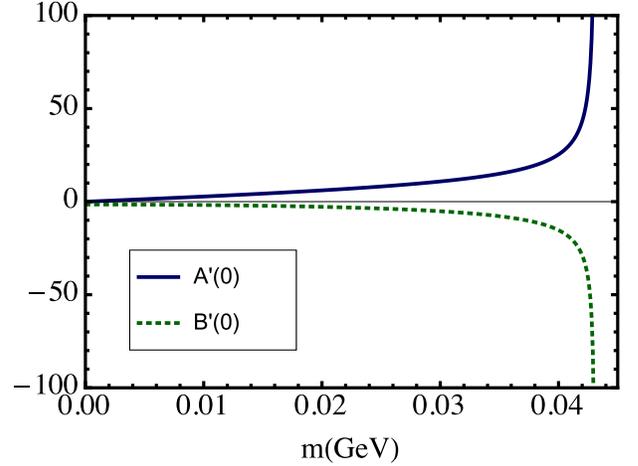}
\caption{The varying of $A^\prime(0)$ and $B^\prime(0)$ with current quark mass.}
\label{fig2}
\end{figure}
Applying the algorithm described in Sec. \ref{dseandalgorithm}, the solutions of gap equation can be obtained with finite current quark mass. In the remain of this paper, we mark the solution deduced by Wigner solution, namely quasi-Wigner solution, with "W" and mark the solution deduced by Nambu solution with "N". The mass functions of quasi-Wigner solution, $M_W(p^2)=B_W(p^2)/A_W(p^2)$, with different current quark masses are displayed in Fig. \ref{fig1}. We can see from Fig. \ref{fig1} that the ultraviolet behavior of the mass functions are consistently tends to their current quark mass as expected. Nevertheless, the infrared limit behavior of the quasi-Wigner solution, $M_W(0)$, keep negative with different current quark masses. As the current quark mass increasing, $M_W(0)$ becomes smaller and smaller.
Our algorithm of solving $A(p^2)$, $B(p^2)$ and $C(p^2)$ depends on the derivative to them, in specific numerical calculations, we find that $B^\prime_W(p^2)$ is divergent at $m=43.1$~MeV. In order to show the details, the $m$ dependence of $A^\prime_W(0)$ and $B^\prime_W(0)$ are displayed in Fig. \ref{fig2}. We can see that $A^\prime_W(0,m)$ and $B^\prime_W(0,m)$ vary very slow for small $m$, but suffer a rapid change around $m=43$~MeV, and finally turn to infinity at $43.1$~MeV. Here it is worth noting that our algorithm could continuously extend the Wigner solution to relatively large current quark mass from chiral limit.

In order to determine which solution is chosen by physics, the comparison between their CJT effective actions is helpful, it's form reads~\cite{Roberts:2000aa}
\begin{equation}
\Gamma_\mathrm{CJT} = \mathrm{Tr} \mathrm{Ln}\left[ S_0^{-1} S \right] + \frac{1}{2} \mathrm{Tr} \left[ 1-S_0^{-1} S \right].
\end{equation}
Each solution of gap equation has an effective action, the nature will choose the lowest value. The difference of CJT effective actions between quasi-Wigner and Nambu solutions are evaluated numerically,
\begin{eqnarray}
\Gamma^W_{\mathrm{CJT}} - \Gamma^N_{\mathrm{CJT}} &=& 8.5\times 10^{-4} \mathrm{GeV}^{4},
\end{eqnarray}
$\Gamma^W_{\mathrm{CJT}}>\Gamma^N_{\mathrm{CJT}}$ implies that the Numbu solution is more stable at $T=0=\mu$. The value, namely vacuum pressure difference, can be deemed as a guidance for the bag constant adopted in studies of compact stars~\cite{PhysRevD.93.065011,Weissenborn:2011qu,PhysRevD.95.056018}.
\subsection{Finite temperature}\label{finiteT}
\begin{figure}
\includegraphics[width=0.5\textwidth]{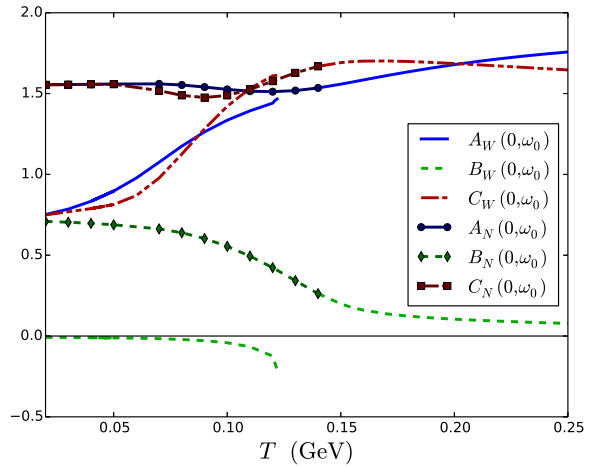}
\caption{$A(0,\omega_0), B(0,\omega_0)$ and $C(0,\omega_0)$ dependence on $T$ for both quasi-Wigner and Nambu modes at $m=5$~MeV.}
\label{fig3}
\end{figure}
The gap equation with finite temperature can be solved by means of the algorithm described in Sec. \ref{dseandalgorithm}. Similar to the case of zero temperature, the behavior of quark propagator tends to free fermion propagator in the ultraviolet limit as expected, and what we actually focus on is the infrared region. The quasi-Wigner mode at $(\vec{p}^2=0, n=0)$, namely the functions $A_W(\vec{p}^2=0,\omega_0)$, $B_W(\vec{p}^2=0,\omega_0)$ and $C_W(\vec{p}^2=0,\omega_0)$, varying with $T$ is plotted in Fig. \ref{fig3}, where current quark mass is chosen as a typical value $m=5$~MeV.
The quasi-Wigner solution is absent in the region of $T \in (122,137)$~MeV because the procedure of the algorithm confronts a singularity when we extend the current quark mass as far as $m=5$~MeV from the chiral limit.
As a comparison, we display the Nambu-Goldstone mode at $(\vec{p}^2=0,\omega_0)$ as well. We can see from Fig.~\ref{fig3} that the Nambu solution ends upto $T=137$~MeV because the Nambu solution disappear in the region of $T>137$~MeV in the chiral limit. $B_N(0,\omega_0)$ monotonously decreases and keeps positive along with $T$ increasing.

\begin{figure}
\includegraphics[width=0.5\textwidth]{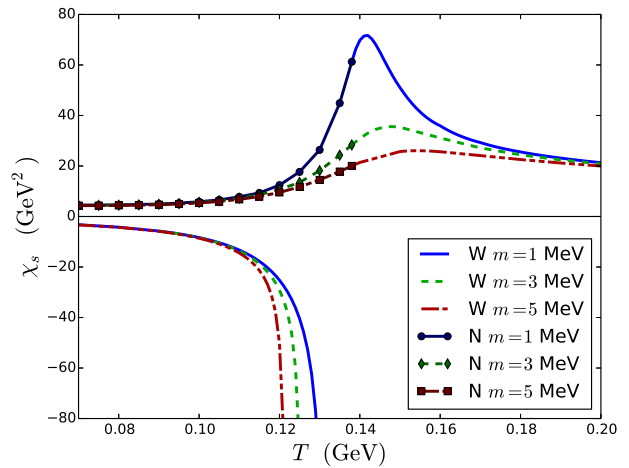}
\caption{The chiral susceptibility $\chi_s$ dependence on T, the "W" represents the susceptibility of quasi-Wigner solution and the "N" means the susceptibility of Nambu-Goldstone solution.}
\label{fig4}
\end{figure}
In order to understand the details of the quasi-Wigner solution, we show the $T$ dependence of its chiral susceptibility~\cite{PhysRevLett.106.172301}, 
\begin{equation}
\chi(0,\omega_0) = \frac{\partial B(0,\omega_0^2)}{\partial m},
\end{equation}
at different current quark masses in Fig. \ref{fig4}. In the chiral limit, both chiral susceptibilities of the Wigner and Nambu solutions have a singularity which located at the same temperature $T=137$~MeV~\cite{PhysRevLett.106.172301}. While in the case of nonzero current quark mass, we can see from Fig.~\ref{fig4} that the susceptibility of quasi-Wigner solution still has a singularity, but the susceptibility of Nambu solution is smoothly varying with $T$. As the current quark mass increasing, the location of singular temperature turns lower and lower. In the area of $T>137$~MeV, only quasi-Wigner solution exist. We can see from Fig.~\ref{fig4} that there is a peak for all $m$, and the height of the peak is decreasing with the increasing of current quark mass. 
In the past, people do not know the multi-solution of gap equation at finite temperature beyond the chiral limit, hence one can not get the difference of effective action of two solutions.
\begin{figure}
\includegraphics[width=0.5\textwidth]{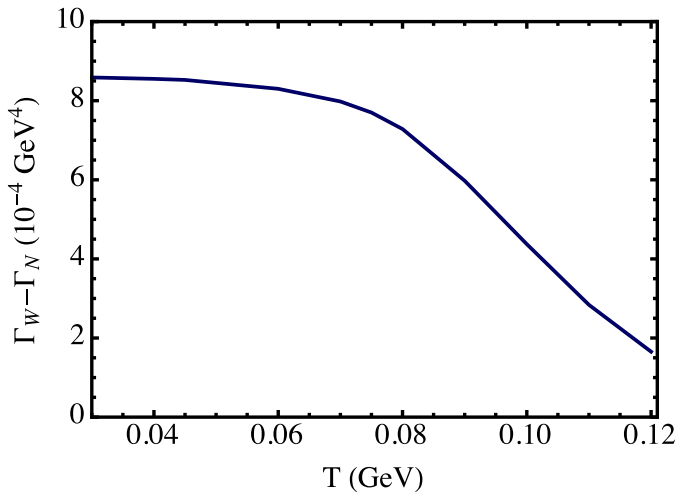}
\caption{The temperature dependence of the difference between the CJT actions of the quasi-Wigner and Nambu solutions.}
\label{fig5}
\end{figure}
Based on our algorithm, the $T$ dependence of the difference between the CJT actions of the quasi-Wigner and Nambu solutions is displayed in Fig.~\ref{fig5}. In the coexistence region of the two solutions, $\Gamma_W-\Gamma_N$ is always positive, which implies that the Nambu solution is more stable when $T<122$~MeV. For $T>122$~MeV, one can not give the difference since there is only one solution.

\subsection{Finite temperature and finite chemical potential}\label{finiteTmu}
\begin{figure}
\includegraphics[width=0.5\textwidth]{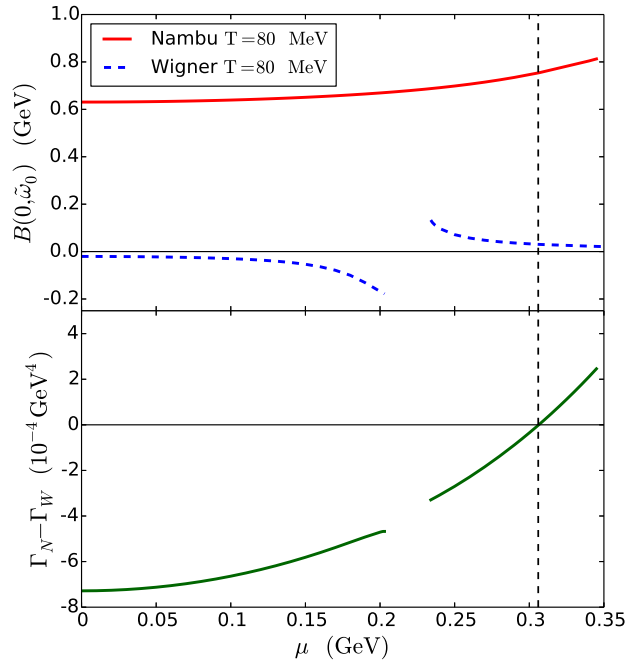}
\caption{Comparison of $B_W(0,\tilde\omega_0)$ and $B_N(0,\tilde\omega_0)$ and the difference of their CJT effective actions.}
\label{fig6}
\end{figure}
The results displayed in Sec. \ref{zeroT} and \ref{finiteT}, indicate that the Nambu solution is always the physical solution in the two solutions coexistence region at vanishing chemical potential. It is expected that the Wigner solution would be physically stable in the chiral symmetry phase if the two solutions coexist.

In order to show the details of the behavior of phase structure, we display $B(0,\tilde\omega_0)$ of  Wigner and Nambu solutions varying with $\mu$ at $T=80$~MeV in Fig. \ref{fig6} as an example. We can see in Fig. \ref{fig6} that $B(0,\tilde\omega_0)$ of Wigner solution is negative at $\mu<203$~MeV. Similar with the case of finite $T$ and zero chemical potential, the Wigner solution can not be solved by the simple iterative method in the region of $203~\mathrm{MeV}<\mu<234$~MeV. From $\mu=234$~MeV, $B(0,\tilde\omega_0)$ turns positive and smoothly decrease with $\mu$. On the other hand, $B_N(0,\tilde\omega_0)$ monotonously increases along with $\mu$. By means of the difference of CJT effective action between Nambu and Wigner solutions, which is showed in Fig. \ref{fig6}, the stable solution can be determined with the vast majority of regions of chemical potential. The Fig. \ref{fig6} shows that the Numbu solution is stable at $\mu<306$~MeV, and the Wigner solution is more stable at $\mu>306$~MeV, the chiral phase transition happens at $\mu=306$~MeV. The property of two phases coexistence is in consistence with the results in NJL model and the chiral limit case of DSEs~\cite{Du:2015psa,Lu:2015naa,PhysRevLett.106.172301}. However, we can see from Fig.~\ref{fig6} that the two solutions coexistence also happens in the area $\mu\in(0,0.2)$~GeV at $T=80$~MeV. It implies that the region of coexistence also exists at low $\mu$.

\begin{figure}
\includegraphics[width=0.45\textwidth]{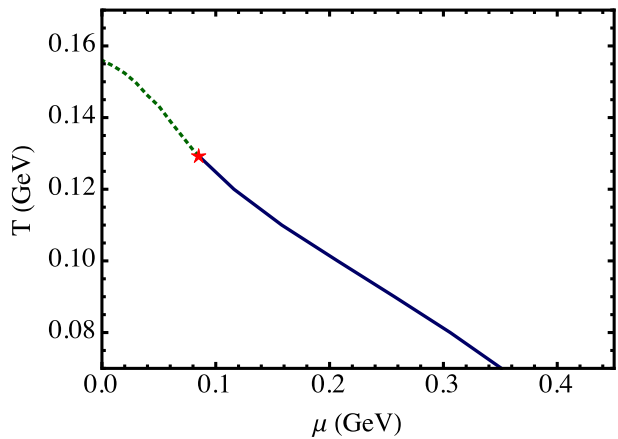}
\caption{The possible QCD phase diagram based on our calculation.}
\label{fig7}
\end{figure}
We performed calculations at many $T$ values, and then plot the phase diagram in Fig.~\ref{fig7}. We find that the first order phase transition occurs at $T<129$~MeV, where two solutions are coexistence. At $(T,\mu)=(129,86)~$MeV, the two solutions coincide and the second order phase transition happens because of the chiral susceptibility is divergent. With in the area $(\mu<86~\mathrm{MeV},T>129~\mathrm{MeV})$, there is only one solution survives, and the chiral susceptibility is smoothly varying with $T$ and $\mu$. The dashed line in Fig.~\ref{fig7} indicates the peak of the chiral susceptibility of the thermal and dense QCD system.
\section{Summary}\label{sum}
We give a briefly introduction to the gap equation at first, and then propose a new algorithm to solve Wigner solution with nonzero current quark mass. Take zero temperature as an example, the key point of this algorithm is that the equations of $\frac{\partial A(p^2,m)}{\partial m}$ and $\frac{\partial B(p^2,m)}{\partial m}$ has unique solution because of their linearity if $A$ and $B$ are solved in advance. Combining with $B(p^2,0)=0$, which is an explicit solution, we obtain the unique quasi-Wigner solution which tightly kindred with the chiral symmetry preserving solution in the chiral limit.

In rainbow truncation and a ultraviolet dumping model, we have studied the quasi-Wigner solution and its susceptibility, the mass functions with different current quark masses are displayed, the far infrared of which decrease along with $m$. The quasi-Wigner solution can be continuously extended upto $m=43.1$~MeV, and thereafter the $\frac{\partial A_W}{\partial m}$ and $\frac{\partial B_W}{\partial m}$ encounters a singularity. 

Further more, we have studied the quasi-Wigner solution at zero and finite chemical potential with finite temperature in rainbow truncation and the Gaussian type model as well. In the case of zero chemical potential, $B_W(0,\omega_0)$ keeps negative and decreases along with $T$ at low temperature, and at medium temperature, namely $T\in(122,137)$~MeV, the quasi-Wigner solution is absent. At relative high temperature, $T>137$~MeV, the quasi-Wigner solution is smoothly connected with Nambu-Goldstone solution. In order to show the details of the quasi-Wigner solution at finite temperature, we display the chiral susceptibility with different current quark masses, there is a divergent point for each $m\leq 5$~MeV, and the singular temperature decreases along with $m$.
In the case of finite chemical potential and temperature, the dependence of $\mu$ of the quasi-Wigner solution is showed at $T=80$~MeV as an example. There exist a critical $\mu_c$ where the CJT effective action equals to that of Nambu solution. Below the $\mu_c$, the system is Nambu phase, and the system turns to quasi-Wigner phase when $\mu>\mu_c$. The behavior of two solutions coexist is disappeared when $T$ goes upto $129$~MeV, and the CEP is located at $(T,\mu)=(129, 86)$~MeV.

\acknowledgments
This work is supported in part by the National Natural Science Foundation of China (under Grants No. 11475085, No. 11535005, and No. 11690030) and national Major state Basic Research and Development of China (2016YE0129300), and 
the China Postdoctoral Science Foundation (under Grants No. 2015M581765 and No. 2016M591809), and the Jiangsu Planned Projects for Postdoctoral Research Funds (under Grant No. 1402006C).


\bibliography{xuss}
\end{document}